\documentclass[%
 reprint,
superscriptaddress,
 amsmath,amssymb,
 aps,
pra,
]{revtex4-2}
\usepackage{amsmath}
\usepackage{graphicx}
\usepackage{dcolumn}
\usepackage{bm}
\usepackage{hyperref}
\hypersetup{colorlinks=true, citecolor=blue, urlcolor=blue, linkcolor=blue}
\usepackage{subfigure}

\begin{document}

\title{Purification of Gaussian States by Photon Subtraction}

\author{Kun Zhang}
\affiliation{Institute of Nonlinear Physics and Department of Physics, Zhejiang Normal University, Jinhua, 321004 Zhejiang, China}
\affiliation{State Key Laboratory of Precision Spectroscopy, Joint Institute of Advanced Science and Technology, School of Physics and Electronic Science, East China Normal University, Shanghai 200062, China}
\author{Huijun Li}
\email{hjli@zjnu.cn}
\affiliation{Institute of Nonlinear Physics and Department of Physics, Zhejiang Normal University, Jinhua, 321004 Zhejiang, China}
\affiliation{State Key Laboratory of Precision Spectroscopy, Joint Institute of Advanced Science and Technology, School of Physics and Electronic Science, East China Normal University, Shanghai 200062, China}
\author{Jietai Jing}
\email{jtjing@phy.ecnu.edu.cn}
\affiliation{State Key Laboratory of Precision Spectroscopy, Joint Institute of Advanced Science and Technology, School of Physics and Electronic Science, East China Normal University, Shanghai 200062, China}
\affiliation{CAS Center for Excellence in Ultra-intense Laser Science, Shanghai 201800, China}
\affiliation{Collaborative Innovation Center of Extreme Optics, Shanxi University, Taiyuan, Shanxi 030006, China}
\author{Nicolas Treps}
\affiliation{Laboratoire Kastler Brossel, Sorbonne Universit\'{e}, CNRS, ENS-Universit\'e PSL, Coll\`{e}ge de France, 4 place Jussieu, F-75252 Paris, France}
\author{Mattia Walschaers}
\email{mattia.walschaers@lkb.upmc.fr}
\affiliation{Laboratoire Kastler Brossel, Sorbonne Universit\'{e}, CNRS, ENS-Universit\'e PSL, Coll\`{e}ge de France, 4 place Jussieu, F-75252 Paris, France}

\begin{abstract}
 Photon subtraction can enhance entanglement, which for pure states induces a decrease in the purity of reduced states.  In contrast, by analyzing the purities of Gaussian states before and after subtracting a single photon, we prove that the purity of a Gaussian state can also be increased by less than 20\%.  On the one hand, it reveals that photon subtraction can reduce entanglement, and on the other hand, it reveals that it can achieve a limited amount of Gaussian state purification.  Through the analysis of some examples, we demonstrate the inherent mechanism and applicable scope of photon-subtraction-based purification. In a multimode system, we find that photon subtraction can increase entanglement and purify some of the reduced states simultaneously. We thus present purification through the suppression of Gaussian noise as a new application for photon subtraction in continuous-variable quantum information processing.   
\end{abstract}

\maketitle

\section{introduction}

Quantum information based on continuous-variable (CV) optics exploits  quadratures of  the electromagnetic fields to encode, transmit, and process quantum information \cite{RevModPhys.77.513}.  One of the key advantages of this approach is that the quantum information processing tasks can be implemented in a deterministic unconditional way  using  easily implemented  Gaussian entangled states and  Gaussian operations \cite{RevModPhys.84.621}.
Several protocols such as quantum teleportation \cite{PhysRevLett.80.869,PhysRevLett.132.100801} and quantum  key distribution \cite{doi:10.1126/sciadv.adi9474,WOS:000259686400018}  have already been demonstrated experimentally.

The noise caused by dissipation and decoherence is the main source of errors in information processing\cite{PhysRevLett.85.1330,PhysRevA.95.062312,Lasota_2023,PhysRevLett.102.130501}. For a long time, there has been a constant pursuit for high-purity or low-noise states to transmit information more faithfully\cite{PhysRevLett.97.053601,PhysRevLett.77.2818,PhysRevLett.76.722}.   
Gaussian operations can effectively suppress non-Gaussian noise by driving non-Gaussian into Gaussian states \cite{PhysRevLett.97.150505,WOS:000261386000010,PhysRevA.75.050302}.  
However, the most common noise, such as fluctuation noise, thermal noise, scattering noise, etc., are all Gaussian, which cannot be suppressed by local Gaussian operations \cite{PhysRevLett.102.120501,PhysRevLett.89.137904,PhysRevLett.89.137903,PhysRevA.66.032316}. For this reason, a method called entanglement distillation is proposed to combat noise by increasing entanglement, rather than directly suppressing noise.  
This  method  aim to extract a small number of high entangled states from a large number of weak entangled states by means of  non-Gaussian operations and classical communications \cite{PhysRevA.53.2046}. 
In addition to the required number of copies of the initial entangled state \cite{PhysRevA.67.062320,PhysRevA.82.042331}, the main characteristic  is the selection of non-Gaussian operations \cite{PhysRevLett.84.4002,PhysRevA.80.032309,PhysRevA.67.022304,PhysRevA.102.012407,PhysRevLett.129.273604}.  

Photon subtraction (PS) is a typical non-Gaussian operation used in entanglement distillation, and has driven breakthrough experimental demonstrations of distillation  protocols due to its practical feasibility in the laboratory \cite{doi:10.1126/science.1122858,PhysRevLett.97.083604, WOS:000275058900021,PhysRevLett.112.070402,PhysRevLett.98.030502}.  
Moreover,  an improved version of the PS-based distillation protocol has been proposed   by adding additional local Gaussian operations,  such as displacement \cite{PhysRevA.86.052339} and squeezing   \cite{PhysRevA.84.062309}. 
On the other hand, the characteristic of PS improving bipartite entanglement has been analyzed through different measurement methods \cite{PhysRevA.108.012411}, such as logarithmic negativity \cite{PhysRevA.84.012335,PhysRevA.65.032314} and von Neumann entropy \cite{PhysRevA.86.012328} for two-mode states, and  Rényi-2 entropy \cite{Zhang2022maximalentanglement}
and purity \cite{PhysRevLett.119.183601,PhysRevLett.121.220501}  for multimode states.  
 However, for mixed states, PS usually reduces the purity of  global entangled states.  In other words, PS-based distillation  does not exert all effort to increase entanglement as it does in pure states, it may allocate a portion of the effort to generate additional noise. This has to make people reconsider  the practical value of PS-based distillation.

 Purity is  a good indicator for evaluating the amount of noise in states. 
 Because both Gaussian and non-Gaussian noise can lead to a decrease in the purity of a state.  Based on purity, this paper studies the feasibility of PS directly combating Gaussian noise.  We reveal  that under certain conditions, PS can increase  the purity of Gaussian states, which means PS can be used to develop PS-based purification protocols. Furthermore, we strictly  prove the upper limit of purity increase for any Gaussian state, which indicates that PS-based purification  has a performance upper limit. 
Through the analysis of single-mode Gaussian states, we reveal the purification mechanism of suppressing noise in one direction from the perspective of phase space.   Through the analysis of an entangled pure state with three modes, we find that PS-based  distillation and PS-based  purification can coexist. These works pave the way for  purifying CV multimode Gaussian states.

The structure of the paper is as follows. In Sec. \ref{II}, we firstly review the general methods for calculating the purity of a CV state, then prove the general condition and upper bound for the increase in purity of any Gaussian state before and after subtracting one photon.  In   Sec. \ref{III}, we demonstrate some examples of PS-based purification scheme.   We will conclude  in  Sec.  \ref{IV}.

\section{Proof of purification upper bound}\label{II}

\subsection{Purity of CV quantum state}
In CV quantum optics, Wigner function is  a representation of an arbitrary quantum state in phase space.  A quantum state $\hat{\rho}^{G}$ that can be represented by a Gaussian Wigner function is called a Gaussian state, which is usually written as \cite{RevModPhys.84.621}
\begin{eqnarray}\label{Wg2}
 W^{G}(\vec \beta) &=& \frac{\mathrm{exp}[{-\frac{1}{2}(\vec\beta-\vec\alpha_0)^tV^{-1}(\vec\beta-\vec\alpha_0)}]}{{(2\pi)^{{m}}\sqrt{\mathrm{det}(V)}}},
\end{eqnarray} 
where $\vec{\beta}=({x}_1, \cdots, {x}_{m},$ ${p}_1, \cdots, {p}_{m})^t\in\mathbb{R}^{2{m}}$  is a set of
amplitude and phase quadrature. ${x}_i $ and ${p}_i$  are the eigenvalues of quadrature operators  $\hat{x}_i $ and $\hat{p}_i$, which have the relation with  the associated annihilation operator $\hat{a}_i$ and  creation operator $\hat{a}_i^{\dag}$ as $\hat{x}_i=\hat{a}_{i}^{\dag}+\hat{a}_{i}$, and   $\hat{p}_{i} = i(\hat{a}^{\dag}_i-\hat{a}_i)$.  $\vec\alpha_0$ and $V$ are  referred to as the displacement vector and  the covariance matrix.  For a Gaussian state,  purity can be directly obtained through the associated covariance matrix by $\mu={1}/{\sqrt{|V|}}$, but it is not true for non-Gaussian states. It is worth noting that all  information about a quantum state   $\hat{\rho}$ can be obtained through  its Wigner function $W_{\mathcal{}}(\vec \beta) $.  For instance, the purity of $\hat{\rho}$ can be obtained by integrating its Wigner function as follows \cite{PhysRevLett.119.183601}
\begin{equation}\label{eq:purityFromWigner}
\mu_{\mathcal{}}=\mathrm{tr}(\hat{\rho}^{2})=(4\pi)^{{m_{\mathcal{}}}}\int_{\mathbb{R}^{2{m_{\mathcal{}}}}}|W_{\mathcal{}}(\vec \beta)|^2 \mathrm{d}^{2{m_{\mathcal{}}}}\beta.
\end{equation}
Similarly,  the variance of the  amplitude quadrature of mode-$k$ can  be obtained by
\begin{eqnarray}\label{eq:varance}
 \Delta^2\hat{x}_k
&=&\int_{\mathbb{R}^{2m}}x_{k}^{2}W_{\mathcal{}}(\beta) \mathrm{d}^{2{m_{\mathcal{}}}}\beta-\left[\int_{\mathbb{R}^{2m}}x_kW_{\mathcal{}}(\beta) \mathrm{d}^{2{m_{\mathcal{}}}}\beta\right]^2.
\end{eqnarray}
This procedure of extracting quantum state information  is still effective for non-Gaussian states.

Here, we are interested in  photon subtracted non-Gaussian state, whose Wigner function can be accurately described through a phase space representation method \cite{PhysRevLett.119.183601,PhysRevA.96.053835}, yielding a comprehensive expression \cite{Zhang2022maximalentanglement}
\begin{eqnarray}\label{wng}
\nonumber W^{-}_{\cal }(\vec\beta_{\cal })=&&[\|{X V^{-1}_{\cal }(\vec\beta_{\cal } - \vec\alpha_{\cal }) - \vec\alpha_g}\|^2 + tr(V_g - X V^{-1}_{\cal } X^t) \\
&& -2 ] \times\frac{ W^{G}_{\cal }(\vec\beta_{\cal })}{\|{\vec\alpha_g}\|^2 + tr(V_g) - 2},
\end{eqnarray}
where the label $g$ indicates the mode in which the photon is subtracted.  
$X = G^t(V-\openone)$, and $\openone$ is the identity matrix.  Here, $G$ is a $2m \times 2$ matrix where the two columns are the basis vectors $\vec g^{(x)}$ and $\vec g^{(p)}$ associated with the two phase space axes of mode $g$:
\begin{equation}
    G = \begin{pmatrix} \mid & \mid \\
    \vec g^{(x)}& \vec g^{(p)}\\
    \mid & \mid
    \end{pmatrix}.
\end{equation}
Analogously, $V_g = G^tVG$, and the displacement vectors $\vec \alpha_{}$ and $\vec \alpha_{g} = G^t\vec \alpha$. We can now use standard techniques of Gaussian integrals to evaluate the purity with Eq. (\ref{eq:purityFromWigner}) from $W^{-}(\vec\beta)$. 
 By comparing the purities of a Gaussian state before and after PS, we can analyze the changes in any Gaussian state. 

\subsection{Upper bound of relative purity}
 Given the vast array of Gaussian states,  it is impractical to investigate all Gaussian states one by one. To derive general conclusions and features, a research methodology that comprehensively encompasses all Gaussian states is required. 
In reference \cite{Zhang2022maximalentanglement}, we have provided a general expression of relative purity 
\begin{eqnarray}\label{RP1}
\nonumber\frac{\mu^{-}}{\mu}=\frac{1}{2}&&+\Bigg[\frac{1}{2}(\sum_{i=1}^{{m}}\frac{\tilde{{N}}_{i}}{n_i})^2+
 \frac{1}{2}|\alpha_{g}|^4+|\sum_{i=1}^{{m}}k_{i}l_{i}\frac{n_i^2-1}{2n_{i}}|^2\\
 \nonumber &&+{\alpha_{g}^{\ast}}^2(\sum_{i=1}^{{m}}\frac{k_{i}l_{i}(n_{i}^{2}-1)}{2n_{i}})+\alpha_{g}^2(\sum_{i=1}^{{m}}\frac{k_{i}^{\ast}l_{i}^{\ast}(n_{i}^{2}-1)}{2n_{i}})\\
&&+|\alpha_{g}|^2\sum_{i=1}^{{m}}N_{i}\Bigg]/{(\sum_{i=1}^{{m}}{N}_{i}+|\alpha_{g}|^2)^2},
\end{eqnarray}
where  $\mu^{\mathcal{}}$ and $\mu^{\mathcal{-}}$ are the purities of the Gaussian state before and after  PS operation, respectively.  ${N}_{i}=|k_{i}|^2\frac{n_i+1}{2}+|l_{i}|^2\frac{n_i-1}{2}$, and ${\tilde{N}}_{i} =|k_{i}|^2\frac{n_i+1}{2}-|l_{i}|^2\frac{n_i-1}{2}$, where $n_i$ is the Gaussian noise factor ($n_i=1$ represents vacuum). And $l_{i}$, $k_{i}$, and  $\alpha_{g}$  are the complex coefficients of the unitary operator $\hat{D}\hat{U}$ transforming operator $\hat{a}_{g}$.  The
specific details are as follows,
\begin{equation}
 \hat{U}^{\dag} \hat{D}^{\dag}\hat{a}_{g} \hat{D}\hat{U} = \alpha_g+\sum_{i}^{m}k_i\hat{a}_{i}^{\dag}+l_i\hat{a}_{i},
\end{equation}
where $\hat{D}$ and $\hat{U}$ come from the thermal decomposition $\hat{\rho}^{G}=\hat{D}\hat{U}\bigotimes\limits_{i=1}^{m}\hat{\rho}_{i}^{th}\hat{U}^{\dag}\hat{D}^{\dag}$ of Gaussian state $\hat{\rho}^{G}$\cite{RevModPhys.84.621}, where $\hat{\rho}_{i}^{th}$ is a single-mode thermal state. According to  the property of the Bogoliubov transformation, $k_i$ and $l_i$ are subject to the following constraint
 \begin{eqnarray} \label{contraint}
\sum\limits_{i=1}^{m}|l_{i}|^2-|k_{i}|^2=1 \quad \mathrm{and} \quad \sum\limits_{i=1}^{m}k_{i}l_{i}^{*}=-\sum\limits_{i=1}^{m}k_{i}^{*}l_{i}. 
          \end{eqnarray}

When ${\alpha}_{g}=0$, it  always holds that ${\mu^{-}}/{\mu}\leqslant{1}$ (See Appendix \ref{appendix1}). This means that for any Gaussian state with ${\alpha}_{g}=0$, PS can only reduce purity, which also means an increase in entanglement entropy, if we consider the initial Gaussian state as a subsystem of a big entangled state. As an irreversible operation, it is natural for PS to exhibit this phenomenon. After some necessary mathematical processing, we have strictly proven that the lower bound of Eq. (\ref{RP1}) is 1/2  \cite{Zhang2022maximalentanglement}. Here,  what we are interested in is whether the upper bound on relative purity is possibly beyond one.

To address this inquiry, we  rewrite the relevant complex parameters as  $\alpha_g=\tilde{\alpha}_g e^{i\varphi}$, $k_i=\tilde{k}_ie^{i\phi_i}$, and $l_i=\tilde{l}_ie^{i\theta_i}$, respectively.  $\tilde{\alpha}_g$, $\varphi$, $\tilde{k}_i$, $\phi_i$,  $\tilde{l}_i$,  and $\theta_i$ are real numbers. Through the derivation in appendix \ref{appendix2}, we obtain the necessary and sufficient conditions for ${\mu^{-}}/{\mu}\geqslant1$ 
 \begin{eqnarray} 
\sum\limits_{i=1}^{m}\frac{\tilde{k}_{i}\tilde{l}_{i}(n_{i}^{2}-1)}{2n_{i}}cos(2\varphi-\phi_i-\theta_i)>0,
          \end{eqnarray}
and
 \begin{eqnarray} \label{condition}
 {\tilde{\alpha}_{g}}^2 &\geqslant & \frac{(\sum\limits_{i=1}^{m}{N}_{i})^2 -(\sum\limits_{i=1}^{m}\frac{\tilde{{N}}_{i}}{n_i})^2-2|\sum\limits_{i=1}^{m}k_{i}l_{i}\frac{n_i^2-1}{2n_{i}}|^2 }{4\sum\limits_{i=1}^{m}\frac{\tilde{k}_{i}\tilde{l}_{i}(n_{i}^{2}-1)}{2n_{i}}cos(2\varphi-\phi_i-\theta_i)}.
          \end{eqnarray}
Eq.(\ref{condition}) takes the equal sign corresponding to ${\mu^{-}}/{\mu}=1$. 
In addition, by making the angle factors satisfy  $2\varphi=2n\pi+\phi_i+\theta_i$  for all $i$,  we also obtain  an upper bound function $f(\alpha)$ that satisfies  ${\mu^{-}}/{\mu} \leqslant f(\alpha)$. This is an reachable bound because the adjustment of the angular  factors does not violate the constraint condition Eq. (\ref{contraint}). 
 $f(\alpha)$  can be expressed as 
    \begin{eqnarray}\label{fa}
 f(\alpha)=1+\frac{1}{2}\frac{[x^2-y^2+\frac{1}{2}z^2+2\alpha z]}{(y+\alpha)^2},
 \end{eqnarray}
where $x=\sum\limits_{i=1}^{m}\frac{\tilde{{N}}_{i}}{n_i}$, $y=\sum\limits_{i=1}^{m}{N}_{i}$, $z=|\sum\limits_{i=1}^{m}2\tilde{k}_{i}\tilde{l}_{i}\frac{n_i^2-1}{2n_{i}}|^2$,  and $\alpha=\tilde{\alpha}_g^2$.
Furthermore, by the derivative of $f(\alpha)$ with respect to $\alpha$,
we obtain the maximum value of  $f(\alpha)$  
  \begin{eqnarray}
  \mathrm{max}(f(\alpha)) =1+\frac{z^2}{2(y^2-x^2+2yz-\frac{1}{2}z^2)}.
  \end{eqnarray}
and its upper open bound 
  \begin{eqnarray}\label{max}
\mathrm{max}( f(\alpha))&<&1+\frac{1}{5+8\frac{\zeta}{1-\zeta}},
  \end{eqnarray}
where $\zeta\in(0,1]$.  The maximum value of right side of Eq. (\ref{max}) asymptotically approaches  $1.2$,  when $\zeta\rightarrow0$. This means that the relative purity always meets ${\mu^{-}}/{\mu}<{1.2}$. That is to say, the purity increased by PS will not reach or exceed 20\% of the original Gaussian state purity.
Therefore, Gaussian states can be
purified by PS and the ability of purification is limited.

\section{Analysis of purification cases}\label{III}

\begin{figure}[h]
\centering
\includegraphics[scale=0.4]{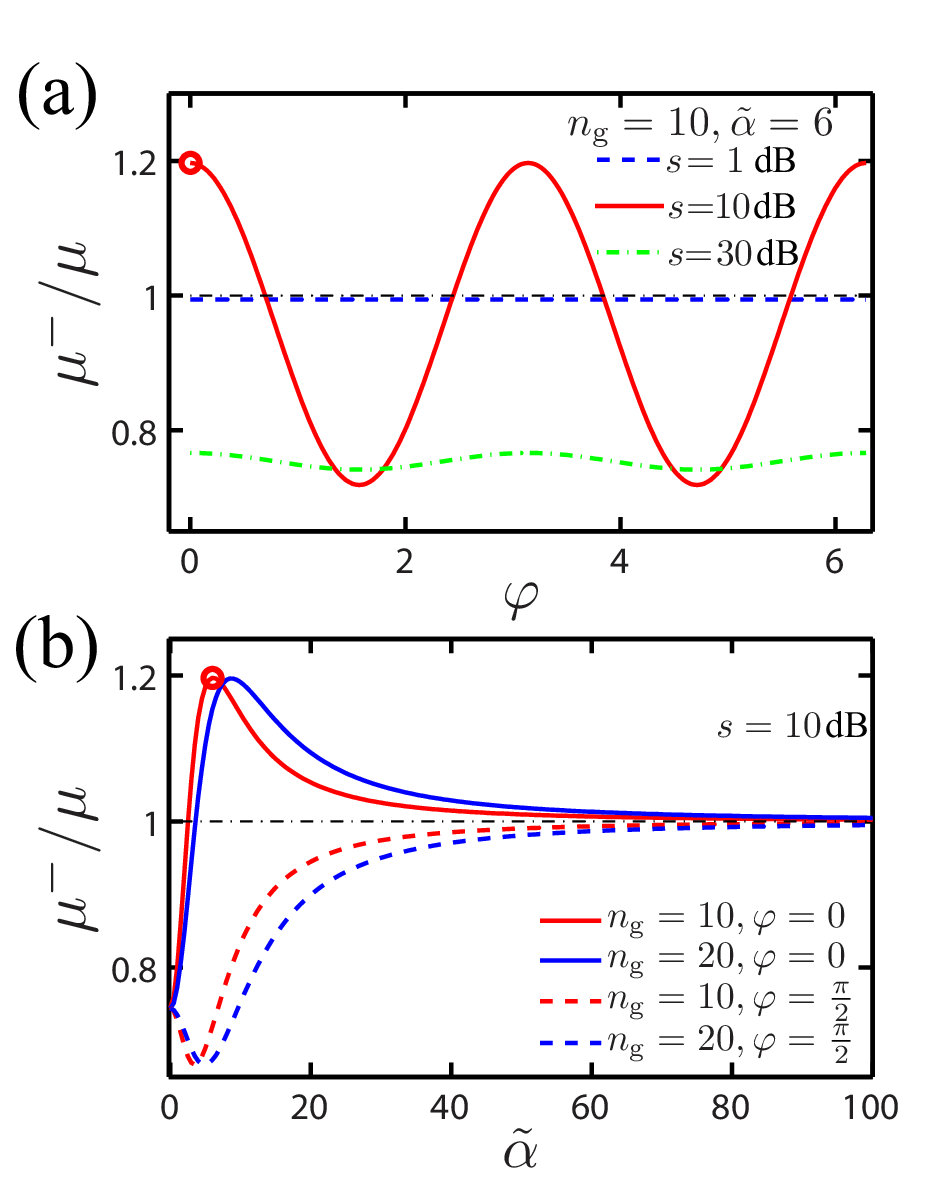}
\caption{\footnotesize (Color online) (a) The curves of relative purity changing with  displacement direction  $\varphi$. The noise and displacement parameters are fixed at $n_{\mathrm{g}}=10$ and $\tilde{\alpha}=6$. The three curves correspond to squeezing parameters of 1dB, 10dB and 30dB, respectively. (b) The curves of relative purity changing with displacement $\tilde{\alpha}$. The squeezing parameter is fixed at $s=10\mathrm{dB}$. The four curves correspond to  four sets of noise and displacement direction parameters,  which are $n_{\mathrm{g}}=10, \varphi=0$; $n_{\mathrm{g}}=20, \varphi=0$; $n_{\mathrm{g}}=10, \varphi=\pi/2$; and $n_{\mathrm{g}}=20, \varphi=\pi/2$. In (a) and (b), all points on the curves that are greater than one  indicate the corresponding Gaussian states can be purified by photon subtraction (PS).  The red circles in (a) represent a set of parameters  $\tilde{\alpha}=6$, $\varphi=0$, $n_g=10$ and $s=10$ that make the relative purity reach the bound $f(\tilde{\alpha})\approx1.1967$. 
} \label{fig0101}
\end{figure}
\begin{figure}[h]
\centering
\includegraphics[scale=0.35]{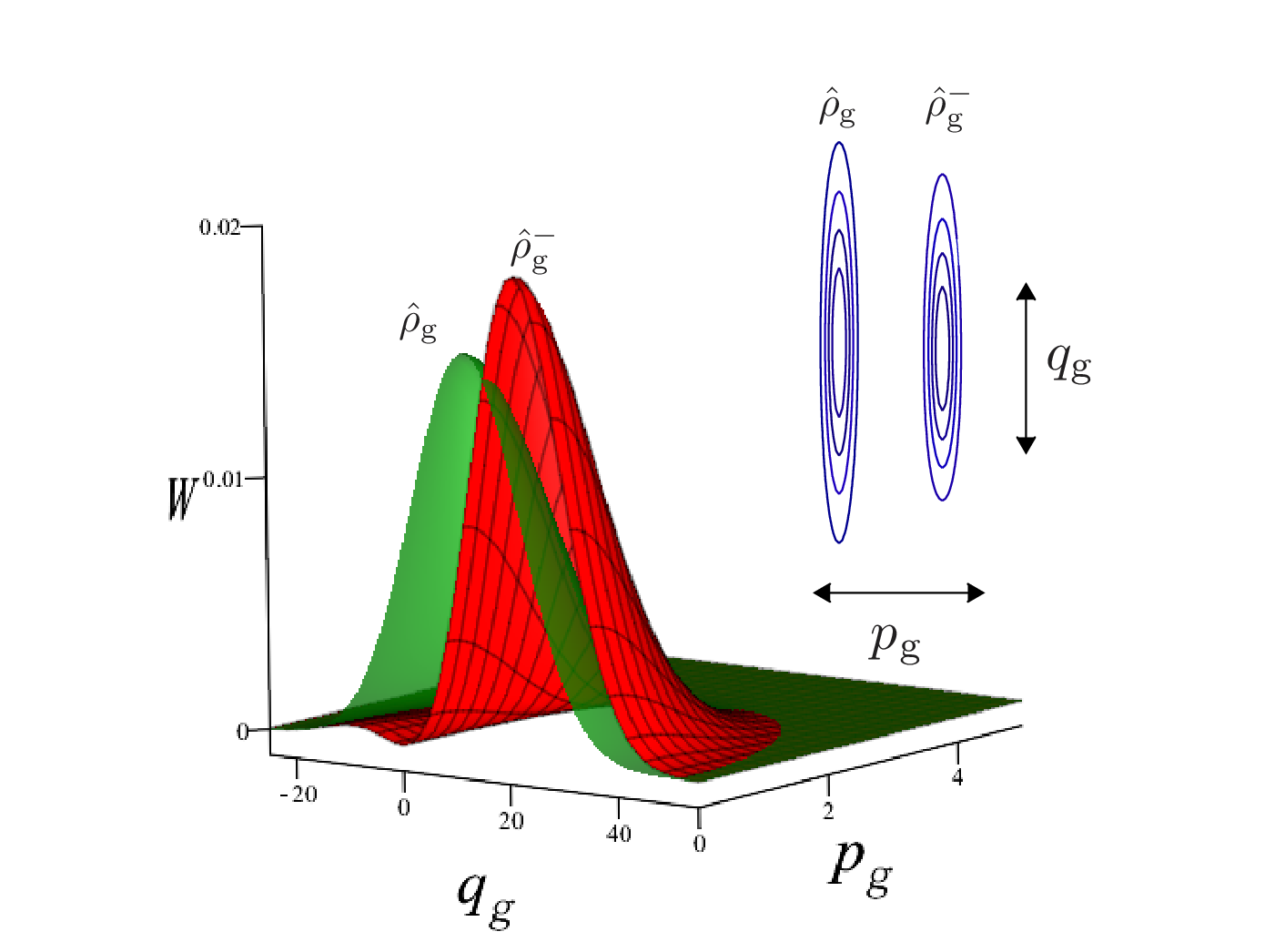}
\caption{\footnotesize (Color online) This graph shows two half Wigner function contours,  with the green one indicating an initial Gaussian state and the red one with a black grid  indicating the non-Gaussian state after subtracting one photon. The relative purity is $\mu^{-}/\mu\approx1.1967$  corresponding to the parameters indicated by the red circle in FIG. \ref{fig0101}(a).  The upper right  corner subgraph is an area chart of the phase space distribution of the two Wigner functions.  From the phase space, it can be intuitively seen that photon subtraction (PS) counteracts Gaussian noise in the anti-squeezing direction and can achieves good purification effect.  
} \label{fig0201}
\end{figure}

\subsection{PS-based Purification of Single-mode Gaussian state} 
We here consider a single-mode state $\hat{\rho}_g $ that can be characterized by a single-mode
covariance matrix  $V_g=\mathrm{diag}(n_gs,n_g/s)$ and a displacement $\alpha_g=\tilde{\alpha}e^{i\varphi}$. 
Due to  this state has only one mode, PS can only be applied to this mode. According to Section \ref{II}, it can be deduced that $\tilde{k}_g=(s-1)/(2\sqrt{s})$ and $\tilde{l}_g=(s+1)/(2\sqrt{s})$ , which means that the angular factor satisfies $\phi_g=\theta_g=0$.  
As shown in FIG. \ref{fig0101},  we draw several curves of relative purity as a function of the parameters $\tilde{\alpha}$, $\varphi$, $n_g$ and $s$. 
 Each point on the curves correspond to a Gaussian state and its photon subtracted state. As shown in FIG. \ref{fig0101}(a), we can always obtain the optimal relative purity $f(\alpha)$ at $\varphi=n\pi$, this result is consistent with   Eq. (\ref{fa}).

As shown by the red dashed curve and the blue dashed curve in  FIG. \ref{fig0101}(b),  when the angle factor takes a mismatched value $\varphi=2\pi$,  no matter how large the displacement distance $\tilde{\alpha}$ is, 
 the purity will never increase.  
 When the squeezing parameter $s=10$ and the matching angle factor $\varphi=0$  are taken, we can always find the maximum value max($f(\alpha)$)  by adjusting the displacement distance $\tilde{\alpha}$,  as  shown by the red circle in  FIG. \ref{fig0101}(b).  
Note that these two real curves share the same maximum value max($f(\alpha)$),  which is close to 1.2. When $s\rightarrow\infty$, $\zeta\rightarrow0$
and then max($f(\alpha)$) can infinitely approaches 1.2.  In short, the range of Gaussian states that can be purified by PS is limited and the degree of purification varies with the initial Gaussian state.

In order to better understand the PS-based purification, 
we  now analyze in detail the case  corresponding to the red circle on the red solid line in   FIG. \ref{fig0101}(a).  According to Eq. (\ref{eq:varance}), the variances of the amplitude and phase quadrature before and after PS have the relation
 \begin{eqnarray}
   {\Delta^{2} \hat{q}_{\mathrm{g}}^{-}}{} &\approx& 0.85\Delta^{2} \hat{q}_{\mathrm{g}}, \quad \mathrm{and}\quad {\Delta^{2} \hat{p}_{\mathrm{g}}^{-}}{} = \Delta^{2} \hat{p}_{\mathrm{g}}.
 \end{eqnarray} 
This shows that PS suppresses noise in  the $q_\mathrm{g}$-direction and has no effect in the $p_\mathrm{g}$-direction. In other words, we reduce the anti-squeezing, while keeping the squeezing unchanged. As shown in FIG. \ref{fig0201}, comparing the two Wigner functions, it shows that PS causes the peak of the Wigner function to become higher and causes a slight displacement of the Wigner function along the $q_\mathrm{g}$-direction, in other words, the suppression of noise is distinct.

\subsection{ PS-based Purification of Multimode  Gaussian state}

\begin{figure}[h]
\centering
\includegraphics[scale=0.4]{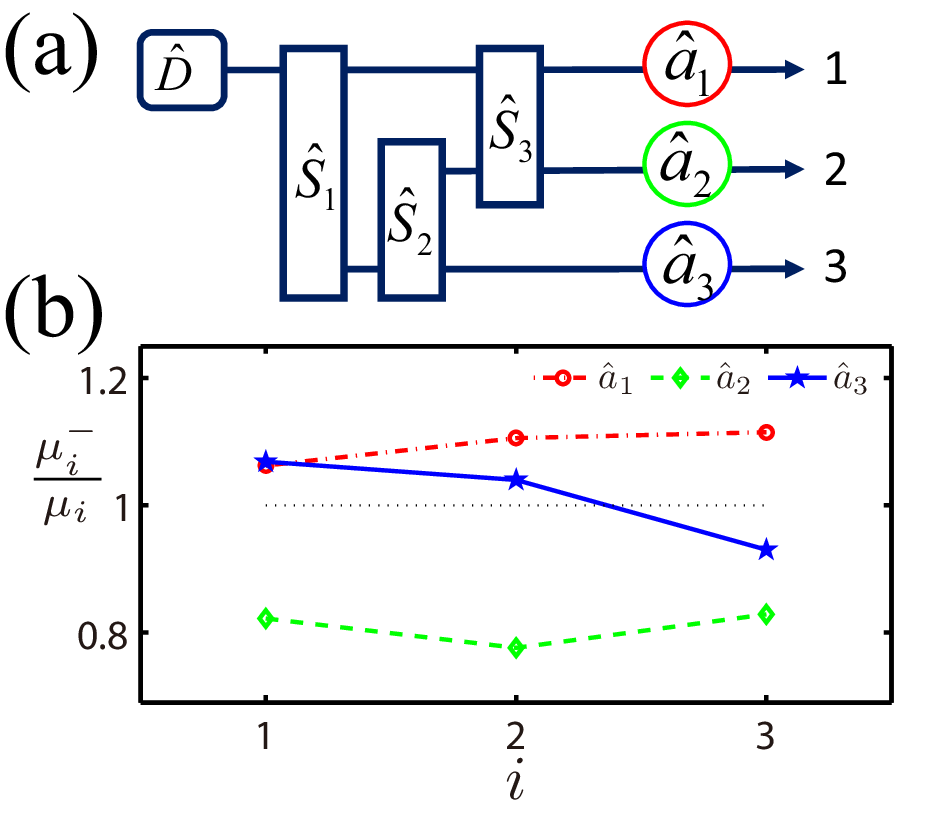}
\caption{\footnotesize (Color online) (a) shows the schematic diagram of generating  an entangled state with three modes.  Firstly, a displacement operator $\hat{D}$ is used to generate a weak coherent light, and then the coherent light is sequentially injected into three two-mode squeezing operators $\hat{S}_1$,$\hat{S}_2$, and $\hat{S}_3$. The displacement parameter is set to $\alpha=1.6$, and the three squeezing  parameters  are all set to $s=3\mathrm{dB}$.  Finally, the three modes are performed photon subtraction (PS) operations, represented by red circle, green circle, and blue circle, respectively. (b) shows the relative purity of each mode under different PS operations.  The small red circles,  green diamonds, and  
blue pentagrams 
represent the relative purities of the three modes when subtracting one photon from mode-$1$, mode-$2$, and mode-$3$, respectively.} \label{fig3}
\end{figure}

When it comes to multimode states,  the number of parameters increases with the number of modes. To achieve purification, besides some methods of changing parameters, directly  changing the mode used to perform PS operation is also a good alternative option.  
As an example,  we consider a three mode Gaussian entangled pure  state, which is expressed as the following state vector 
\begin{eqnarray}\label{threeGaussianstate}
|\psi\rangle_{123}=\hat{S}_3\hat{S}_2\hat{S}_1\hat{D}|0,0,0\rangle,
  \end{eqnarray}
which can be formed by connecting three vacuum states using one displacement operation $\hat{D}$ and three two-mode squeezing  operations $\hat{S}_i(i=1,2,3)$, as shown in FIG. \ref{fig3}(a).  We arrange a displacement operation $\hat{D}$ here so that the three modes can generate a small displacement. This is necessary because without displacement, photon subtraction will not increase the purity of any state, which has been strictly proven in  Appendix \ref{appendix1}. 
We perform PS operation  on the three modes separately, as  shown by the red, green, and blue circles  in  FIG. \ref{fig3}(a). Then  we calculate the relative purity of each mode under each PS operation and plot it in   FIG. \ref{fig3}(b).  Multimode systems have a replicated entanglement structure. In terms of bipartite entanglement, a system with three modes has three types of bipartite entanglement, which are the entanglements between the three mode and their remaining subsystems. Since the global state $|\psi\rangle_{123}$ is a pure state, and PS preserves this global purity, the increase (decrease) in the purity of a mode directly indicates a decrease (an increase) in bipartite entanglement between the mode and the remaining subsystem. 

When one photon is subtracted from mode-$1$, the relative purities of the three modes are all greater than one, as shown by the red circles in  FIG. \ref{fig3}(b). This means that the entanglement of all bipartite entangled structures is reduced.  On the contrary, when one photon is subtracted from mode-$2$, the relative purities of the three modes are all less than one, as shown by the green diamonds in  FIG. \ref{fig3}(b). This means that the entanglement of all bipartite entangled structures increases. When we subtract one photon from mode-3, the purities of both mode-1 and mode-2 increase, while the purity of mode 3 decreases, as shown by the blue pentagrams in  FIG. \ref{fig3}(b). This implies that the bipartite entanglement between mode-3 and its subsystems increases, while  the bipartite  entanglement between mode-1 or mode-2 and their respective remaining subsystems decreases.  
Interestingly, as the entanglement between mode-3 and its remaining subsystems increases, the purity of mode-1 and mode-2 in the remaining subsystem  increases. This means that PS can purify the modes of a reduced subsystem while increasing the entanglement of the global system.  This means that PS-based distillation and  purification can be achieved simultaneously in one system.

\section{conclusion}\label{IV}
This paper finds that PS can be used to develop  purification protocols for Gaussian states, and studies the properties of such protocols.  Firstly, 
based on  the general expression of relative purity in Eq. (\ref{RP1}), we  prove that the purity of a Gaussian state can be increased by less than 20\%, and provide the sufficient and necessary conditions for purity increase.  These results directly reflect the  performance and implementation conditions of PS-based purification protocols.  Then,  based on the analysis of single-mode Gaussian states,  we use phase space to visually demonstrate the physical mechanism of PS-based purification. That is to suppress Gaussian noise along a single direction in phase space. Finally, with the help of a three-mode entangled state, we demonstrate a strategy of achieving different purification effects by changing the photon subtracted mode. And we find  that PS-base distillation and  purification can coexist in one system. 

In general, for a pure bipartite entangled system, without considering the threshold point ${\mu_{}^{-}}/{\mu_{}}=1$, PS either purifies or distills it, which leads to either reducing or increasing the entropy of the subsystem. 
This paper reveals a special case that lies between pure purification and pure distillation, 
in which the entanglement of an global bipartite  system can be enhanced, while the noise of some modes in the subsystems can be suppressed.  
 In a sense, this paper demonstrates the flexibility and plasticity of PS in multimode systems, although whether impure systems can have similar properties 
 is still  an open question.

\section{ACKNOWLEDGMENT}
K.Z. and J.J. acknowledge financial
support through the National Natural Science Foundation of China (12225404, 12434014, 11874155); Innovation Program of Shanghai Municipal Education Commission (2021-01-07-00-08-E00100); Program of Shanghai Academic Research Leader (22XD1400700); Natural Science Foundation of Shanghai (17ZR1442900); Minhang Leading Talents (201971); the 111 project (B12024).
H.L. acknowledge  financial support through the 
Project supported by Natural Science Foundationof Zhejiang Province of China(LZ22A050002), National Natural Science Foundation of China(12074343).
M.W. and N.T. acknowledge financial support from the ANR JCJC project NoRdiC (ANR-21-CE47-0005), the Plan France 2030 through the project OQuLus (ANR-22-PETQ-0013), and the QuantERA II project SPARQL that has received funding from the European Union’s Horizon 2020 research and innovation programme under Grant Agreement No 101017733.

  \appendix
\section{ When $\alpha=0$} \label{appendix1}
Let $\alpha=0$,  Eq. (\ref{RP1}) becomes
\begin{eqnarray}\label{RP}
\frac{\mu_{}^{-}}{\mu_{}}
 =&&\frac{1}{2}+\frac{\frac{1}{2}(\sum\limits_{i=1}^{m}\frac{\tilde{{N}}_{i}}{n_i})^2+|\sum\limits_{i=1}^{m}k_{i}l_{i}\frac{n_i^2-1}{2n_{i}}|^2}{(\sum\limits_{i=1}^{m}{N}_{i})^2}.
\end{eqnarray}
where ${N}_{i} =|k_{i}|^2\frac{n_i+1}{2}+|l_{i}|^2\frac{n_i-1}{2}$ and ${\tilde{N}}_{i} =|k_{i}|^2\frac{n_i+1}{2}-|l_{i}|^2\frac{n_i-1}{2}$.
For any complex number $\alpha$, it always holds that $(\alpha+\alpha^{\ast})^2\geqslant0$ and $(\alpha-\alpha^{\ast})^2\leqslant0$, thus   the value range of $\alpha^2+{\alpha_{\mathrm{g}}^{\ast}}^2$  is
\begin{eqnarray}\label{alpha2}
-2|\alpha_{\mathrm{g}}|\leqslant \alpha_{\mathrm{g}}^2+{\alpha_{\mathrm{g}}^{\ast}}^2\leqslant2|\alpha_{\mathrm{g}}|^2.
\end{eqnarray}
Then we get
\begin{eqnarray}\label{e3}
  |\sum_{i=1}^{m}\frac{k_{i}l_{i}(n_{i}^{2}-1)}{2n_{i}}|^2 &\leqslant&
(\sum_{i=1}^{m} \sqrt{|k_{i}|^2|l_{i}|^2}\frac{(n_{i}^{2}-1)}{2n_{i}} )^2.
\end{eqnarray}
Bring Eq. (\ref{e3}) into Eq. (\ref{RP})
\begin{eqnarray}
 \nonumber \frac{\mu_{}^{-}}{\mu_{}}
 &\leqslant&\frac{1}{2}+\frac{\frac{1}{2}(\sum\limits_{i=1}^{m}\frac{\tilde{{N}}_{i}}{n_i})^2+(\sum\limits_{i=1}^{m} \sqrt{|k_{i}|^2|l_{i}|^2}\frac{(n_{i}^{2}-1)}{2n_{i}} )^2}{(\sum\limits_{i=1}^{m}{N}_{i})^2}\\
 &=&\frac{1}{2}+\frac{1}{2}\frac{(\sum\limits_{i=1}^{m}\frac{\tilde{{N}}_{i}}{n_i})^2+\frac{1}{2}(\sum\limits_{i=1}^{m} \sqrt{|k_{i}|^2|l_{i}|^2}\frac{(n_{i}^{2}-1)}{n_{i}} )^2}{(\sum\limits_{i=1}^{m}{N}_{i})^2}.
\end{eqnarray}
We know that $\frac{|k_{i}|^2+|l_{i}|^2}{2}\geqslant \sqrt{|k_{i}|^2|l_{i}|^2}$, then
\begin{eqnarray}\label{e12}
  \frac{(\sum\limits_{i=1}^{m} \sqrt{|k_{i}|^2|l_{i}|^2}\frac{n_{i}^{2}-1}{n_{i}} )^2}{2}
 \leqslant  \frac{(\sum\limits_{i=1}^{m} (|k_{i}|^2+|l_{i}|^2)\frac{n_{i}^{2}-1}{n_{i}}) ^2}{8}.
\end{eqnarray}
Due to  the constraints of Eq. (\ref{contraint})  and $(|k_{i}|^2+|l_{i}|^2)\frac{1}{n_{i}}\leqslant (|k_{i}|^2+|l_{i}|^2)n_{i}$,
the right side of Eq. (\ref{e12}) is  less than or equal to zero. This means that 
\begin{eqnarray}\label{c28}
\frac{(\sum\limits_{i=1}^{m}\frac{\tilde{{N}}_{i}}{n_i})^2+\frac{1}{2}(\sum\limits_{i=1}^{m} \sqrt{|k_{i}|^2|l_{i}|^2}\frac{(n_{i}^{2}-1)}{n_{i}} )^2}{(\sum\limits_{i=1}^{m}{N}_{i})^2} \leqslant  1,
\end{eqnarray}
 which means that  it always holds that $\frac{\mu_{}^{-}}{\mu_{}}\leqslant 1$. 

\section{ When $\alpha\neq0$}\label{appendix2}
Now we rewrite Eq. (\ref{RP1}) as follows
\begin{eqnarray}\label{RP2}
\nonumber\frac{\mu^{-}}{\mu}=1&&+\frac{1}{2}\Bigg[(\sum_{i=1}^{{m}}\frac{\tilde{{N}}_{i}}{n_i})^2+
 \frac{1}{2}|\sum_{i=1}^{{m}}2k_{i}l_{i}\frac{n_i^2-1}{2n_{i}}|^2\\
 \nonumber &&+{2\alpha_{g}^{\ast}}^2(\sum_{i=1}^{{m}}\frac{k_{i}l_{i}(n_{i}^{2}-1)}{2n_{i}})+2\alpha_{g}^2(\sum_{i=1}^{{m}}\frac{k_{i}^{\ast}l_{i}^{\ast}(n_{i}^{2}-1)}{2n_{i}})\\
 &&-(\sum_{i=1}^{{m}}N_{i})^2\Bigg]/{(\sum_{i=1}^{{m}}{N}_{i}+|\alpha_{g}|^2)^2}.
\end{eqnarray}
It is not difficult to find that only the second term on the right side of Eq. (\ref{RP2})  is greater than or equal to zero to ensure that ${\mu_{}^{-}}/{\mu_{}}\geqslant1$, that is,   
 \begin{eqnarray} \label{B2}
 \nonumber &&2{\alpha_{\mathrm{g}}}^2(\sum_{i=1}^{m}\frac{k_{i}^{\ast}l_{i}^{\ast}(n_{i}^{2}-1)}{2n_{i}})+2{\alpha^{\ast}_{\mathrm{g}}}^2(\sum_{i=1}^{m}\frac{k_{i}l_{i}(n_{i}^{2}-1)}{2n_{i}})\\&&+(\sum\limits_{i=1}^{m}\frac{\tilde{{N}}_{i}}{n_i})^2+\frac{1}{2}|\sum\limits_{i=1}^{m}2k_{i}l_{i}\frac{n_i^2-1}{2n_{i}}|^2\geqslant (\sum\limits_{i=1}^{m}{N}_{i})^2. 
          \end{eqnarray}      
For convenience, the parameters can be  rewritten as $\alpha_g=\tilde{\alpha}e^{i\varphi}$, $k_i=\tilde{k}_ie^{i\phi_i}$, and $l_i=\tilde{l}_ie^{i\theta_i}$, where we  set $\tilde{k}_i\geqslant 0$,  $\tilde{l}_i\geqslant 0$, and $\tilde{\alpha}\geqslant 0$. The complex angle factors can control the positive or negative signs of these parameters, which do not affect the generality of Eq. (\ref{RP2}).  According to Eq. (\ref{alpha2}),
when taking $2\varphi=2n\pi+\phi_i+\theta_i$  for all $i$, it can be found that $(\sum\limits_{i=1}^{m}\frac{\tilde{{N}}_{i}}{n_i})^2$ and $(\sum\limits_{i=1}^{m}{N}_{i})^2$ remain unchanged, while $2{\alpha_{\mathrm{g}}}^2(\sum\limits_{i=1}^{m}\frac{k_{i}^{\ast}l_{i}^{\ast}(n_{i}^{2}-1)}{2n_{i}})+2{\alpha^{\ast}_{\mathrm{g}}}^2(\sum\limits_{i=1}^{m}\frac{k_{i}l_{i}(n_{i}^{2}-1)}{2n_{i}})$ and $\frac{1}{2}|\sum\limits_{i=1}^{m}2k_{i}l_{i}\frac{n_i^2-1}{2n_{i}}|^2$ can reach  their maximum values.  Thus, an  optimal value of ${\mu_{}^{-}}/{\mu_{}}$ can be obtained as 
 \begin{eqnarray}\label{f}
 \frac{\mu_{}^{-}}{\mu_{}}&\leqslant&1+\frac{1}{2}\frac{[x^2-y^2+\frac{1}{2}z^2+2\alpha z]}{(y+\alpha)^2}
  =f(\alpha),
 \end{eqnarray}
where $x=\sum\limits_{i=1}^{m}\frac{\tilde{{N}}_{i}}{n_i}$, $y=\sum\limits_{i=1}^{m}{N}_{i}$, $z=|\sum\limits_{i=1}^{m}2\tilde{k}_{i}\tilde{l}_{i}\frac{n_i^2-1}{2n_{i}}|^2$,  and $\alpha={\tilde{\alpha}}^2$.  $f(\alpha)$ can be understood as the reachable  boundary function  of ${\mu_{}^{-}}/{\mu_{}}$. Because for any ${\mu_{}^{-}}/{\mu_{}}$, we can always obtain an optimal value by adjusting the complex angle factor. It is theoretically feasible because it does not violate the constraints of Eq. (\ref{contraint}).

Thus, the derivative of $f(\alpha)$ with respect to $\alpha$ is
 \begin{eqnarray}
f(\alpha)^{\prime}=\frac{\mathrm{d}f(\alpha)}{\mathrm{d}\alpha} &=& \frac{  y^2-x^2+yz-\frac{1}{2}z^2-z\alpha}{(y+\alpha)^3}.
 \end{eqnarray}
When $\alpha=\frac{y^2-x^2+yz-\frac{1}{2}z^2}{z}$, we get that
  \begin{eqnarray}
f(\alpha)^{\prime}=0, \quad \mathrm{and} \quad  f(\alpha)^{\prime\prime}<0.
 \end{eqnarray}
 So the maximum value of  $f(\alpha)$ is
  \begin{eqnarray}
  \mathrm{max}(f(\alpha)) =1+\frac{z^2}{2(y^2-x^2+2yz-\frac{1}{2}z^2)}.
  \end{eqnarray}
Due to $y\geqslant x>0$, $z>0$, and $y-x>z$, $\mathrm{max}(f(\alpha))$ follows the following constraint 
  \begin{eqnarray}\label{Bmax}
  \mathrm{max}(f(\alpha))<1+\frac{1}{5+8\frac{x}{y-x}}.
  \end{eqnarray}  
Due to  $y\geqslant x>0$, we can introduce a variable $\zeta$ such that  $x=\zeta y$, where $\zeta\in(0,1]$. Due to  $x>0$ , $\zeta$  cannot reach zero.  
Thus Eq. (\ref{Bmax}) becomes  
  \begin{eqnarray}
\mathrm{max}( f(\alpha))&<&1+\frac{1}{5+8\frac{\zeta}{1-\zeta}}
=f(\zeta)
  \end{eqnarray}
Due to $f(\zeta)^{\prime}<0$,  the maximum value of $f(\zeta)$ asymptotically approaches  $1.2$,  when $\zeta\rightarrow0$. 
This means that any relative purity must satisfy  ${\mu^{-}}/{\mu}<{1.2}$.

\bibliography{apssamp}

\end{document}